\documentclass{aa}
\usepackage{graphics}
\begin{document}

\thesaurus{03 ( 11.01.2; 11.05.1; 11.10.1; 11.14.1; 
13.18.1 )}
\title{The Fanaroff-Riley Transition and the Optical
Luminosity of the Host Elliptical Galaxy}

\author{Gopal-Krishna\inst{1} 
         \and Paul J.\ Wiita\inst{2} }

\offprints{Gopal-Krishna}

\institute{
National Centre for Radio Astrophysics, Tata Institute of Fundamental
Research, 
Pune University Campus, Post Bag No.\ 3, Ganeshkhind, Pune 411 007,
India 
(krishna@ncra.tifr.res.in)
\and
Department of Astrophysical Sciences, Princeton University,
Princeton, NJ 08544-1001, USA (wiita@astro.princeton.edu);\\
on leave from the Department of Physics \& Astronomy, Georgia State University,
 University Plaza,
Atlanta, GA 30303-3083, USA  (wiita@chara.gsu.edu)}

\date{Received 14 November 2000 / Accepted 9 April 2001}

\authorrunning{Gopal-Krishna \& Wiita}
\titlerunning{FR transition and luminosity of the host galaxy}
\maketitle

\begin{abstract}
We show that a model for radio source dynamics we had earlier
proposed can readily reproduce the relationship
between the radio power division separating the two Fanaroff-Riley
classes of extragalactic radio sources and the optical luminosity
of the host galaxy, as found by Owen and Ledlow (1994). 
 In our scenario,
when less powerful jets eventually slow down to the point that the
advance of the working surface (i.e., hotspot) becomes subsonic 
with respect to
the external gas, the jet's collimation is severely weakened.  This
criterion distinguishes the powerful and well collimated
 FR II sources from the weaker sources producing 
the less collimated FR I type morphologies.

\end{abstract}

\section{Introduction}

A landmark in the study of extragalactic 
radio sources was the demonstration by Fanaroff and Riley (\cite{fanaroff})
of the existence of a  relatively sharp morphological transition 
at a radio luminosity of $P^*_R$ corresponding to
$P_{\rm 178 MHz} \simeq 2.5 \times 10^{26}h^{-2}_{50}$ W Hz$^{-1}$. 
The great majority of sources below this 
luminosity (FR I type) are characterized by having diffuse
radio lobes, with their brightest regions within the inner
half of the radio source, and so can be called edge-dimmed.
On the contrary, more powerful sources are usually straighter,
exhibit edge-brightened (FR II)
morphology, and typically contain hotspots near the outer
edges of their radio lobes.

More recently it was realized that the critical radio luminosity
separating the FRI and FR II
actually increases with the optical luminosity
of the host elliptical galaxy, so that $P^*_R \propto L_{\rm opt}^{1.65}$
(as measured from Fig.\ 1 of Ledlow \& Owen \cite{ledlow}); also see
Owen \& White (\cite{owen91}); Owen \& Ledlow (\cite{owen94})
for earlier indications of this effect.  Although many detailed differences
between the host galaxies of the FR I and FR II sources have been
discovered (e.g., Baum et al.\ \cite{baum}; Zirbel 
\cite{zirbel}; for a recent summary,
see Gopal-Krishna \& Wiita \cite{gkw00}), it is fair to state that a key  fact
is that the more luminous the host galaxy is, 
the more powerful the radio source must be in order to attain the FR II 
morphological classification.

The most widely known explanation for the difference between
the FR I and FR II sources is that, while the jets in  both cases
start out moving at very high (relativistic) speeds, those in FR II sources
remain that way out
to multi-kpc distances, while those in the FR I's decelerate to
much lower speeds within a few kpc of the core (e.g., Begelman 
\cite{begelman};
Bicknell \cite{bicknell84}; De Young \cite{deyoung}; Komissarov
\cite{komissarov}; Laing \cite{laing93}, \cite{laing96};
Laing et al.\ \cite{laingetal99}).
Detailed  models for decelerating relativistic jets were
developed by Bicknell (\cite{bicknell94}, \cite{bicknell95}).  The core of his argument is
that relativistic jets, once they  come into pressure
equilibrium with the external interstellar medium (ISM)
or intracluster medium (ICM), will become strongly
Kelvin-Helmholtz unstable and then entrain substantial
amounts of cold ambient material. Bicknell (\cite{bicknell95}) showed that this 
instability
typically set in as the internal jet Mach number approaches $\sim 2$, 
corresponding to internal bulk velocities having dropped to
about $0.6 c$.  

Of course, many other explanations for the FR I/FR II dichotomy
have been proposed (e.g., Baum et al.\ \cite{baum}; Reynolds et al.\  
\cite{reynoldsa}, 
\cite{reynoldsb}; Meier \cite{meier}; Valtonen \& Hein\"am\"aki \cite{valtonen};
 see the recent discussion by Gopal-Krishna \&
Wiita \cite{gkw00}).
Some of these proposals involve fundamental differences 
in the nature of the jets (electron/positron vs.\ electron/proton
plasma) or of the central engine (black hole
spin, type of accretion disk).  The viability of any member of
this class of explanations is
challenged by the existence of HYbrid MOrphology Radio Sources,
or HYMORS, which have a clear
FR I
morphology on one side of the host galaxy, but a distinct
FR II morphology 
on the other side (Gopal-Krishna \& Wiita (\cite{gkw00}, \cite{gkw01}).

An alternative approach posits that FR I and FR II sources
differ primarily in the importance of the beam thrust relative to the basic
parameters of the ambient medium 
(Gopal-Krishna \& Wiita \cite{gkw88} (GKW88); 
Gopal-Krishna  \cite{gk91} (GK91); 
Gopal-Krishna et al.\ \cite{gkwh} (GKWH96);
Blandford 1996).  
In this version of the deceleration scenario, the emphasis 
is on the slowing of the advance of
the hotspot, or working surface, at the end of the jet,
rather than on the slowing of the bulk flow within the jet (e.g., Bicknell
\cite{bicknell95}).
When the hotspot's advance becomes transonic relative
to the ambient medium, its Mach disk weakens considerably
due to the fall in ram pressure,
and the jet becomes decollimated; this soon leads to an FR I
morphology.  Further expansion of such jets beyond this
point is expected to be in the form of a plume, as discussed 
in GKW88, where
it was also shown that for reasonable values of jet and galaxy
parameters (density, core radius and temperature of
the ISM) such a jet flaring can occur within several
kpc of the core for jet powers below about $10^{43}$ erg s$^{-1}$,
consistent with the observations of the FR~I/FR~II
division (GKW88; GK91; GKWH96).  Also, in this picture, the 
concomitant dimunition/cessation of the `backflow' of the beam
plasma would cause a depletion of the protective sheath of beam plasma 
around the jet, facilitating entrainment of the ISM material into
the jet flow.

Bicknell (\cite{bicknell95}) expanded his model 
to account for the Owen--Ledlow transition in the $P_R$--$L_{\rm opt}$
plane.  A key ingredient in his scenario involved connecting the 
jet dynamics to empirically established  relations between the optical
magnitude of the galaxy on the one hand, and the soft X-ray luminosity,
core radius, and central velocity dispersion of elliptical galaxies
on the other hand.  In the present study, we follow Bicknell and adopt 
these same empirical relationships.
Bicknell then ties the galactic parameters to those of the
jet by demanding that the establishment of a pressure balance
between the jet and the external medium corresponds closely to 
the location of the jet's 
internal transonic transition.  By doing so, he was able to 
derive a formula connecting beam power to the optical luminosity
of the host galaxy.  In this paper we follow a different approach
in this latter stage.

Additional assumptions on the efficiency of the conversion of the 
jet's energy into total radio luminosity ($> 1-2\%$),
and thence into monochromatic radio emission, allowed 
Bicknell (\cite{bicknell95}) to obtain a fairly good fit
to the slope of the Owen--Ledlow division, finding that $P^*_R \propto L_{\rm
opt}^{2.1}$, as well as getting a decent fit to the intercept; however, there were
quite a few poorly constrained parameters in his model.
Nevertheless, Bicknell (\cite{bicknell95}) argued that the the slope of the relation was
rather insensitive to the likely uncertainties in parameters.
He further argued that the
intercept would tend to be driven towards better agreement
with the data for  plausible variations in those
parameters, which included: the ratio of the jet's lifetime to the
synchrotron cooling time of the highest energy electrons;
the ratio of the upper and lower cutoff energies for
the electron distribution;
the ratio of jet radius to its length at the transition distance;
the ratio of the product of the jet pressure and square of its
radius evaluated at the transition radius to that of the external
medium evaluated at its core radius; the bulk velocity of the
jet at the transition point; the low frequency radio spectral index;
and the value of an integral (discussed below) which depends on
an upper cut-off radius for the X-ray emitting halo of the host
galaxy.  

 The new magnetic switch model (Meier \cite{meier})
fundamentally distinguishes FR I from FR II sources through the
different speeds of rotation of the magnetic field lines of
their central engines, which are tied to the different spin
rates of their supermassive black holes.  While this scenario can 
also produce a
slope for the $P^*_R - L_{\rm opt}$ dividing line close to that found by
Ledlow \& Owen (\cite{ledlow}) (Meier \cite{meier}) we recall that it is hard
to reconcile this scheme
with the existence of HYMORS (Gopal-Krishna \& Wiita \cite{gkw00}, \cite{gkw01}).
Venturing beyond the standard nuclear jet paradigm, 
gravitational slingshot models can in principle be compatible with
the Owen--Ledlow relation (Valtonen \& Hein\"am\"aki \cite{valtonen}), and
could also account for HYMORS (Gopal-Krishna \& Wiita \cite{gkw00}).

\section{The Model}

In an earlier study involving ``weak headed quasars'', where
prominent one-sided jets are not seen to terminate in a conspicuous hotspot,
we have argued that dissipation of jet power, as suggested
by Swarup et al.\ (\cite{swarup}) and Saikia et al.\ (\cite{saikia}), may actually not be
responsible for the lack of terminal hotspots (GKWH96).
Taking a clue from the fact that no two-sided jets are seen in these sources, 
we argued that the lack of a hotspot could be best explained through the
 onset of a jet's decollimation when the hotspot's (or, nearly
equivalently, the bow shock's)
velocity becomes transonic relative to the external medium (GKWH96).
This successful explanation of this class of radio
sources motivates us to seek an explanation for the Owen--Ledlow
relation in terms of a similar scenario.  
This model makes no assumptions about the relativistic nature
of the bulk velocity of the internal
jet fluid, which may be gradually decelerating as the head of the jet advances.

Our study makes use of the same empirical relations between
the elliptical's blue magnitude, $M_B$, and its soft X-ray emission,
$L_{\rm X}$ (Donnelly, 
Faber and O'Connell \cite{donnelly}),
stellar velocity dispersion, $\sigma$ (Terlevich et al.\ \cite{terlevich} --- the 
Faber--Jackson relation),
 and X-ray core radius, $a$ (Kormendy \cite{kormendy}),
as employed by
Bicknell (\cite{bicknell95}), assuming $H_0 = 75$ km s$^{-1}$ Mpc$^{-1}$:
\begin{equation}
{\rm log}~ L_{\rm X} = 22.3 - 0.872 M_{\rm B}, 
\end{equation}
\begin{equation}
{\rm log}~ \sigma = 5.412 - 0.0959 M_{\rm B},   
\end{equation}
\begin{equation}
{\rm log}~ a = 11.7 - 0.436 M_{\rm B}.
\end{equation}

However, we adopt the initially conical jet model based on our
earlier work (Gopal-Krishna \& Wiita \cite{gkw87}), which was generalized in 
Gopal-Krishna et al.\ (\cite{gkws89}, GKWS89) to
allow for bulk relativistic
jet flow whereby the hotspot motion can also
be relativistic.  Ram pressure balance,
under these conditions, gives the hotspot velocity, $v$, as
a function of distance, $D$, from the central engine, as (GKWS89):
\begin{equation}
v(D) = {\frac {X c[1+(D/a)^2]^{\delta/2}} {D + X[1+(D/a)^2]^{\delta/2}}}.
\end{equation}
Here $X = (4 L_b/\pi c^3 \theta^2 n_o m_p \mu)^{1/2}$,
where $L_b$ is the jet (beam) power, $\theta$ is the
jet's effective opening angle, $m_p$ and $\mu$ have their
usual definitions,
and $n_0$ is the central density of the ISM, which is
taken to fall off as (e.g., Forman et al.\ \cite{forman}; Canizares et
al.\ \cite{canizares}; Conway \cite{conway})
\begin{equation}
n(D) = {\frac {n_0}{[1+(D/a)^2]^{\delta}}}.
\end{equation}
For our calculations we choose
$\theta = 0.1$ rad, which is certainly justified as appropriate
for the inner jet regions, 
particularly since it is now recognized
that the jet's thrust acts on a larger area than the instantaneous
hot spot (e.g., Scheuer's \cite{scheuer} `dentist-drill', as supported by many
three-dimensional numerical simulations: Norman \cite{norman}; Clarke \cite{clarke};
 Kaiser \&
Alexander \cite{kaiser}; Hooda \& Wiita \cite{hooda98}).  
Following Bicknell (\cite{bicknell95}),
we further assume that the X-ray emitting
ISM gas temperature is tied to the central stellar velocity dispersion,
$\sigma$, via,
$kT = 2.2 \mu m_p \sigma^2 /\delta$,  which is Fall's (\cite{fall}) relation between
circular velocity and $\sigma$.  Here
$\delta$ is roughly 0.75, as inferred from soft X-ray images
of nearby ellipticals (e.g., Forman et al.\ \cite{forman}; 
Canizares et al.\ \cite{canizares}; Sarazin
\cite{sarazin}), as well as dynamical models of radio sources
(e.g., Gopal-Krishna \& Wiita \cite{gkw91}; Conway \cite{conway}).

Studies of many sub-classes of young radio sources,
including Compact Symmetric Objects (e.g., Owsianik \& Conway \cite{owsianik};
Conway \cite{conway}),
Gigahertz Peaked Sources 
(Carvalho \cite{carvalho}) and
Compact Steep Spectrum radio sources (Gopal-Krishna \& Wiita \cite{gkw91};
Jeyakumar et al. \cite{jeyakumaretal}) have indicated that the mean density of the
gas interacting with the radio lobes in the
inner $\sim 1$ kpc or so is typically a few atoms cm$^{-3}$, or
several times that estimated from X-ray emission.  This evidence
in favor of a significant contribution of cooler gas to the ambient
medium (at least in the inner portions of elliptical galaxies)
is further supported by studies of the linear-size distribution
of these radio sources (O'Dea \& Baum \cite{odea}).
We shall parameterize the relation between the confining central 
density, $n_0$, used in
Eq.\ (5) and that derived from X-ray measurements, $n_X$ as follows:
$n_0 = \kappa n_X$, with $\kappa \ge 1$; the fiducial value we use
below is $\kappa = 3$.

The radio data cited above as well as the relation between
$M_{\rm B}$ and the X-ray core radius (Eq.\ 3) lead to values of
of $a \sim 1$ kpc.  Therefore
the ISM density begins to approach the
large scale behavior, and thus declines quite rapidly,
at radial distances beyond a few kpc.  Hence, the most likely
regime for the jets' decollimation due to the hot spots
having slowed to subsonic speeds lies within roughly 10 kpc
of the core.  If the hot spot manages to retain supersonic
speed out to such radial distances, then in most
cases the jet is likely to propagate ahead down the
rapidly declining ISM density, preserving its FR II character.
In some cases another possibility for jet flaring 
can arise farther out when the jet crosses
 the pressure-matched interface between the ISM and the
intergalactic medium (GKW88; Wiita et al.\ \cite{wiita90};
 Wiita \& Norman \cite{wiita92}; Hooda et al.\ \cite{hoodaetal}; Hooda \& Wiita
\cite{hooda96}, \cite{hooda98};
Zhang et al.\ \cite{zhang}).   Relatively few sources will, however, make a
transition to FR I morphology upon crossing this
interface since a significant degree of beam collimation 
appears to occur by such radial distances within the extended halos of
most galaxies (Blundell et al.\ \cite{blundelletal}; Jeyakumar \& Saikia 
\cite{jeyakumar}).  
For such very extended sources, our model is clearly
oversimplified, as the  evolution appears to break from self-similarity
and the decrease in $P_R$ with time (even for constant $L_b$)
cannot be ignored (e.g., Blundell \& Rawlings \cite{blundell}).

Other types of sources that may not be readily accommodated in our
simple picture are, for instance, cases where a narrow jet is seen to terminate in a 
bright knot far away (say $> 50$ kpc) from the nucleus, and flares only beyond that point
(e.g., 3C 130,  Hardcastle 1998). Of course, higher values of $D^*$ are expected for sources 
for which the beam's  power and opening angle lie near the upper and lower ends of their
respective ranges.  Also, the possibility of a bright knot and jet expansion occurring 
when the jet interacts with irregularities
in the ambient medium must be considered (e.g., Wang et al.\ \cite{wang}). In this context it is 
especially interesting to recall that recent 
radio observations of the nearest radio galaxy, Centaurus A, have provided clear evidence that a radio 
jet can even emerge recollimated after flaring upon hitting a gas 
cloud (Morganti et al.\ \cite{morganti}). These Australian Telescope Compact Array 
observations have revealed that the flaring of the northern jet of Cen A, a few kiloparsecs 
away from from the nucleus, gives rise to the inner radio `lobe', out of which a narrow
jet is seen to emerge and propagate further out for another few kiloparsecs before flaring 
and fading away in the fashion of typical  FR I jets. Morganti et al. (\cite{morganti}) have interpreted this 
striking `leaky bulb' type morphology as being a manifestation of the  de Laval nozzle 
model proposed by Blandford \& Rees (\cite{br}).

In our analytical formulation, we evaluate 
the critical beam power, $L^*_b$, for which
the hot-spot deceleration to subsonic velocities occurs
at a fiducial distances, $D^*$, of 3 and 10 kpc from the core, for a range of
absolute magnitudes of the host galaxy between $M_{\rm B} = -19$ and $-23.5$.
The 10 kpc value is a typical distance at which jets
flare in a sample of radio galaxies (O'Donoghue et al.\ \cite{odonoghue}).
A recent study of a sample of 38 FR I sources by Laing et al.\ (\cite{laingetal99}) 
gives a mean projected value of $\sim 3.5$ kpc for the radial distance
of the point where the kiloparsec scale jet first becomes visible,
after an initial emission gap (using our value of $H_0$). 
Our main results are not very sensitive to the
choice of $D^*$, at least until the very most luminous galaxies
are considered, and the predicted slope of the separation
 between FR I and FR II sources that
we derive below is not strongly dependent upon it.

We now write the relevant expressions as follows:
\begin{equation}
X = C_2 n_0^{-1/2} L_b^{1/2};
\end{equation}
where $C_2 = (4/ \pi c^3 \theta^2 m_p \mu)^{1/2} = 2.14 \times 10^{-3}$ for
$\mu = 0.620$.
We also require,
$C_1 = 4\pi (0.909) \Lambda(T) I_X$,
where the cooling function, $\Lambda(T)$ is
a rather weak function of the temperature; assuming $T = 10^7$K,
$\Lambda(T) \simeq 1.17 \times 10^{-23}$
(Bicknell \cite{bicknell95}), and the integral over the gas density
distribution gives 
$I_X \simeq 5.0$ (for $\delta = 0.75$), a value we adopt,
following Bicknell (\cite{bicknell95}).  With these choices,
$C_1 = 6.68 \times 10^{-22}.$

Equations (1--3) and (6) can now be combined to yield:
\begin{equation} 
{\rm log}~L_{\rm X} = {\rm log}~C_1 + 2 {\rm log}~n_0 + 3 {\rm log}~a,
\end{equation}
\begin{equation}
{\rm log}~n_X = \log n_0 - \log \kappa = 4.188 + 0.218 M_{\rm B},
\end{equation}
\begin{equation}
{\rm log}~X = -0.109 M_{\rm B} + {\rm log}~C_2 + 0.25 {\rm log}~C_1 + 3.20 + 
0.5 {\rm log}~L_b.
\end{equation}
We next impose the transonic condition on the LHS of Eq.\ (4), via
\begin{equation}
v(D) = c_s = [\gamma 2.2 /\delta]^{1/2} \sigma.
\end{equation}
Taking the ratio of specific heats for the combined plasma,
$\gamma = 5/3$, and using $\delta = 0.75$, this becomes
\begin{equation}
{\rm log}~v(D) = 0.345 + {\rm log}~\sigma = -4.720 - 0.0959 M_{\rm B}.
\end{equation}

Equating this to the RHS of (4), treated  as a function  of $X$, 
gives,
\begin{equation}
{\rm dex}(-4.720 -0.0959 M_{\rm B}) = {\frac {X} {D/[1+(D/a)^2]^{\delta/2} + X}},
\end{equation}
as the relation that must be solved for $L^*_b$ in terms of
$M_{\rm B}$.

\section{Results and Conclusions}

The values of $L^*_b$, $a$, and $n_0$ can be found for specified
values of $M_{\rm B}$ and $D^*$ by
solving Eqs.\ (3), (8), (9) and (12).  The results  are given in Table 1,
along with, $\beta$, the local slope of the $\log L^*_b$--$\log L_{\rm opt}$
 relation.
For the choice of $D^* = 10$ kpc, these
slopes, $\beta$, vary little for $M_{\rm B} > -23.5$, and correspond
closely to a relation  where $L^*_b \propto L_{\rm opt}^{1.6}$;
for $D^* = 3$ kpc, $\beta$ is only slightly less steep.
For the most luminous galaxies, the value of $\beta$ from the
model flattens somewhat.  

In the absence of a definitive approach to compute the 
efficiency, $\epsilon$,
of conversion of the beam power into synchrotron radiation,
we make the commonly adopted first order approximation that the monochromatic 
radio luminosity is a constant fraction of the beam power 
(at least near the transition luminosity).
Then the local slope remains the same, 
with $P^*_{\rm R, 1.4 GHz} \stackrel{\sim}{\propto} L_{\rm opt}^{1.56}$,
and the key features of the Owen--Ledlow diagram
are satisfactorily explained, with a very good fit to the
slope retained.  The intercept also fitted well, provided the product
$\epsilon \kappa \simeq 0.27$ for $D^* = 10$ kpc.  As shown in Fig.\ 1, this 
is in excellent agreement with the bifurcation between FR I
and FR II radio sources, and nominally
provides even a better fit than does Bicknell's (\cite{bicknell95}) model, which
yields a power-law index above 2.0.  We use a mean color index for
elliptical galaxies
$(B-R) = 1.15$ taken from a CCD photometric study of cluster galaxies
(J{\o}rgensen et al. \cite{jorgensen}) 
to convert from the absolute blue magnitudes used in Eqs.\ (1--3)
to the red magnitudes plotted by Ledlow et al.\ (\cite{ledlow00});
the dependence of color index on optical magnitude  is small over this
range and we have not attempted to correct for it
(J{\o}rgensen et al. \cite{jorgensen}; Kodama et al.\ \cite{kodama};
Peletier et al. \cite{peletier}).

We note that earlier estimates based on the application of beam models
for FR II sources are in reasonable agreement with such
values of $\epsilon \approx 0.1$ (Gopal-Krishna \& Saripalli \cite{gks}; Dreher
\cite{dreher}; Saripalli
\& Gopal-Krishna \cite{saripalli}),
particularly if a ``ready to radiate'' electron-positron jet composition is
assumed; and evidence for such a composition is
growing (e.g., Reynolds et
al.\ \cite{reynoldsa}; Kaiser \& Alexander \cite{kaiser}; Kaiser et
al. \cite{kaiseretal}; 
Hirotani et al.\ \cite{hirotani}).  
Given that our analysis involves a fairly
large set of empirical relations for elliptical galaxies,
the agreement of our prediction with the observational data is
quite encouraging (Fig.\ 1).

It is perhaps worth remarking on the consistency of our model with
the recent Very Long Baseline Interferometry measurements for the expansion of very small 
and young Compact Symmetric Object (CSO)
radio sources (e.g., Owsianik \& Conway \cite{owsianik} (OC)).  For the
CSO 0710$+$439, the best values obtained by OC are
$V \simeq 0.13 h^{-1} c$ at a distance $D \simeq 25 h^{-1}$pc ($\ll a$),
for $n_0 \simeq 2$ cm$^{-3}$, and $L_b \simeq 5 \times 10^{44}$erg s$^{-1}$.
Inserting the last three values into Eq.\ (4), and choosing our
nominal $\theta = 0.1$ rad,
we obtain $V = 0.18 h^{-1} c$, in quite reasonable agreement;
a somewhat higher value for $\theta$ might be inferred from the maps in OC,
thereby reducing our computed velocity and improving the agreement.  Less
well determined values of expansion speeds obtained by other groups
for 3 other CSO radio sources quoted by OC range from
$0.07 - 0.13 h^{-1} c$ and would easily be accommodated by more typical values
of $L_b$.  The  derived range, $0.08 < \epsilon < 0.31$, for 0710$+$439 (OC)
is also in good agreement with our model.

The forgoing analysis is admittedly oversimplified; 
for instance, we have considered
a fixed opening angle for beams of all 
powers, and have also assumed specific values of $\delta$ and $\kappa$,
all of which are certain to vary somewhat from source to source.  
On the other hand, the division between FR I and FR II sources is
not perfect by any means, and such a spread in properties would allow
for the small number of sources found on the `wrong' side of the dividing line
in Fig.\ 1.
Also, we have 
ignored any variation in the jet
disruption length with the galaxy's optical
luminosity by typically choosing $D^* =$ 10 kpc; 
however, such a variation is quite
plausible, and therefore we have also provided results for $D^* =$ 3 kpc
in the last two columns in Table 1.  

To obtain essentially as good a fit with $D^* =$ 3 kpc, one would have
to raise $\epsilon$ to 0.16 if $\kappa = 3$ instead of $\epsilon = 0.09$ for $D^* =$ 10 kpc; of course
both values could drop if $\kappa > 3$ is considered.
As a first step towards
a more realistic model,  one could consider a distribution of 2--10 kpc
for $D^*$ over the range in the hosts' optical
magnitudes, taking smaller $D^*$ for lower $L_{\rm opt}$.
Then the 
predicted slope of the $P^*_R - L_{\rm opt}$ relation clearly would
be slightly steeper (cf.\ columns 4 and 6 of Table 1), further 
improving the accord with the
data presented by Ledlow \& Owen (\cite{ledlow}) and recently
updated by Ledlow et al.\ (\cite{ledlow00}).  We also note that
the exact results quoted assume a specific spectral index ($\alpha = 1.0$)
and upper and lower cut-offs to the radio band (10 MHz and 100 GHz)
respectively, in relating the monochromatic radio power to the beam power.  
Assuming a flatter overall spectrum, say $\alpha = 0.7$, would not affect
the slope of the relation but would demand that the 
product $\epsilon \kappa$ rise to $\simeq 0.52$ (for $D^* =$ 10 kpc) 
to normalize it.
Despite these uncertainties, it is probably fair to note that
overall our variant model employs several fewer
parameters than does Bicknell's (\cite{bicknell95}), although in
connecting optical and X-ray properties we have closely followed his
approach.

The available very detailed 
maps of a few kiloparsec-scale FR I jets provide evidence for the 
existence of a slower moving sheath of synchrotron plasma 
surrounding a relativistic spine
(e.g., Laing \cite{laing93}, \cite{laing96}; Laing et al.\  \cite{laingetal99}).  
In Bicknell's picture
this sheath arises from turbulent mixing of the
ambient plasma after the jet's Mach number becomes low
enough to allow an exponential growth of the planar Kelvin-Helmholtz 
instability.  In our picture, the
entrainment process is accelerated once the hotspot's
motion has become subsonic,  as this results in the
cessation of the `backflow' of the jet plasma;
this leads to the diminution  of the protective cocoon 
around the  jet core which had hitherto separated the latter from the 
ISM material.  We also recall that the impingement of the
`backflow' onto the jet engenders some of the reconfinement shocks
(e.g. Norman et al.\ \cite{normanetal}; Hooda \& Wiita \cite{hooda96},
\cite{hooda98}) so that
the depletion of the shrouding cocoon plasma reduces their effects
and therefore contributes to the jet's decollimation.
Thus, in our picture, the dramatic slowing down of the beam flow can be thought
of as an eventual outcome of the decollimation of the beam's head
(leading to FR I structure), instead of being its principal cause.

\begin{acknowledgements} 
We thank Michael Ledlow and Frazer Owen for providing the data 
used in Fig.\ 1 and B.\ Premkumar for assistance in producing that
figure.
GK appreciates hospitality at Princeton University while much
of this work was carried out.  PJW acknowledges
support from CST funds at Princeton and RPE funds at Georgia State University.
\end{acknowledgements}
%\clearpage

%\newpage
\begin{figure}
\resizebox{\hsize}{!}{\includegraphics{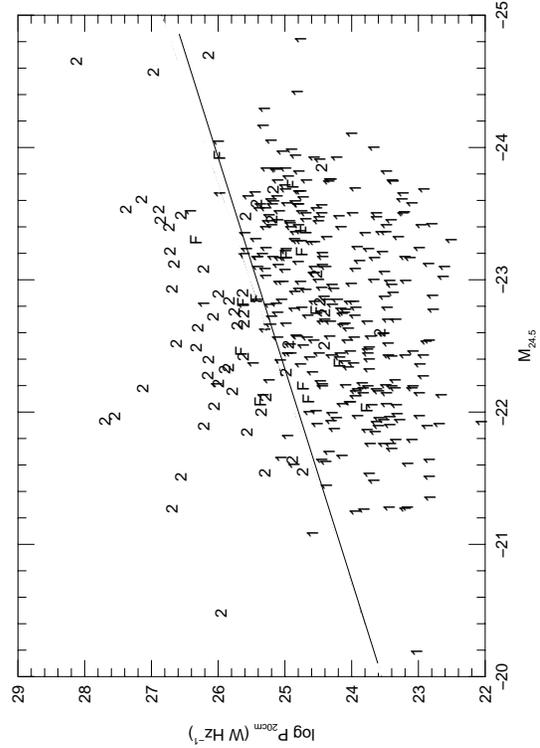}}
% I have been unable to rotate and simultaneously resize this file
\caption{Radio flux plotted against host galaxy magnitude,
with the numerals indicating FR class;
data taken from Ledlow et al.\ (2000).   
 The superimposed curve is the  division between FR I and FR II 
sources arising from our model, taking
$D^*  = 10$kpc and a constant fraction of total beam power
converted into synchrotron emission of $\epsilon = 0.09$.}
\end{figure}

%\newpage

\begin{table}
\caption{Galaxy and jet parameters}
\begin{tabular}{ccccccc}

$M_{B}$ &  $a$& $n_0$  & 
$L_{b, D^*=10}$ & $\beta_{D^*=10}$ &
$L_{b, D^*=3}$ &  $\beta_{D^*=3}$\\
&(kpc)&(cm$^{-3}$)& (erg s$^{-1}$)&& (erg s$^{-1}$)\\

%$-$19.0&0.031&1.112&6.48(40)&    &3.52(40)&    \\ Uncorrected
$-$19.5&0.052&2.82&4.35(41)&1.74&2.42(41)&1.80\\
$-$20.0&0.085&2.02&8.28(41)&1.40&4.53(41)&1.36\\
$-$20.5&0.141&1.57&1.71(42)&1.57&7.67(41)&1.58\\
$-$21.0&0.233&1.22&3.57(42)&1.60&1.92(42)&1.56\\
$-$21.5&0.384&0.951&7.23(42)&1.56&3.93(42)&1.55\\
$-$22.0&0.635&0.741&1.49(43)&1.58&7.92(42)&1.52\\
$-$22.5&1.05&0.576&3.06(43)&1.55&1.56(43)&1.47\\
$-$23.0&1.73&0.447&6.21(43)&1.55&2.80(43)&1.27\\
$-$23.5&2.86&0.348&1.23(44)&1.47&4.11(43)&0.99\\
%$-$24.0&4.73&0.090&7.73(43)&1.39&1.93(43)&0.59\\ Uncorrected
%$-$24.5&7.81&0.070&1.30(44)&1.10&2.18(43)&0.26\\ Uncorrected
%$-$25.0&12.9&0.055&1.84(44)&0.75&2.26(43)&0.08\\ Uncorrected
\end{tabular}
\end{table}

%%%%%%%%%%%%%%%%%%%%%%%%%%%%%%%%%%%%%%%%%%%%%%%%%%%%%%%%%%%%%%%%%%%%%%%%%

\begin{thebibliography}{}


\bibitem[1995]{baum} Baum, S.A., Zirbel, E.L.,  \& O'Dea, C.P. 1995, ApJ,  451, 88

\bibitem[1982]{begelman} Begelman, M.C.  1982, in  IAU Symp.\ 97: Extragalactic Radio
Sources, ed.\ D.S.\ Heeschen  \& C.M.\ Wade
 (Kluwer, Dordrecht),  223

\bibitem[1984]{bicknell84} Bicknell, G.V. 1984, ApJ,  286, 68

\bibitem[1994]{bicknell94} Bicknell, G.V. 1994, ApJ,  422, 542

\bibitem[1995]{bicknell95} Bicknell, G.V. 1995, ApJS, 101, 29

\bibitem[1996]{blandford}  Blandford, R.D. 1996, in 
 Cygnus A -- Study of a Radio Galaxy, ed.\ C.\ Carilli \& D.\ Harris
(Cambridge Univ.\ Press, Cambridge),  264

\bibitem[1974]{br} Blandford, R.D., \& Rees, M.J. 1974, MNRAS, 169, 395

\bibitem[2000]{blundell} Blundell, K.M., \& Rawlings, S. 2000, ApJ, 119, 1111

\bibitem[2001]{blundelletal} Blundell, K.M., Rawlings, S., Willott, C.J., Kassim, N.E.,
\& Perley, R.A. 2001, New Astron. Rev., Proc.\ STScI Workshop on Lifecycles 
of Radio Galaxies, in press

\bibitem[1987]{canizares} Canizares, C.R., Fabbiano, G., \& Trinchieri, G.
1987, ApJ, 312, 503

\bibitem[1999]{carvalho} Carvalho, J.C. 1998, A\&A, 329, 845


\bibitem[1996]{clarke}  Clarke, D.A. 1996, in Energy Transport in Radio
Galaxies and Quasars, ASP Conf.\ Ser., Vol.\ 100,
ed.\ P.\ Hardee, A.H.\ Bridle, \& J. Zensus
(ASP, San Francisco),   311

\bibitem[2001]{conway} Conway, J. 2001, New Astron. Rev., Proc.\ STScI Workshop 
on Lifecycles  of Radio Galaxies, in press

\bibitem[1993]{deyoung}  De Young, D.S. 1993, ApJ,  405, L13

\bibitem[1990]{donnelly}  Donnelly, R.H., Faber, S.M., \& O'Connell, R.M. 1990,
ApJ, 354, 52

\bibitem[1984]{dreher}  Dreher, J.W., 1984, in Physics of Energy Transport in 
Extragalactic Radio Sources, ed.\ A.H.\ Bridle \& J.A.\ Eilek  
(NRAO, Green Bank),   109

%\bibitem[1989]{eilek}  Eilek, J.A., \&  Shore, S. 1989, ApJ, 342, 18

\bibitem[1987]{fall}  Fall, S.M. 1987, in 
Nearly Normal Galaxies from the Planck Time
to the Present, ed.\ S.M.\ Faber (Springer Verlag, New York),   326

\bibitem[1974]{fanaroff}  Fanaroff, B.L.,  \& Riley, J.M. 1974, MNRAS,  167, 31P 

\bibitem[1985]{forman}  Forman, W., Jones, C., \& Tucker, W.
1985, ApJ, 293, 102

\bibitem[1991]{gk91}  Gopal-Krishna  1991, A\&A,  248, 415 (GK91)

\bibitem[1984]{gks}  Gopal-Krishna, \& Saripalli, L. 1984, A\&A, 139, L19

\bibitem[1987]{gkw87}  Gopal-Krishna, \&  Wiita, P.J. 1987, MNRAS, 226, 531

\bibitem[1988]{gkw88}  Gopal-Krishna, \&  Wiita, P.J. 1988, Nature 333, 49 (GKW88)

\bibitem[1991]{gkw91}   Gopal-Krishna, \&  Wiita, P.J. 1991, ApJ,  373, 325

\bibitem[2000]{gkw00}   Gopal-Krishna, \& Wiita, P.J. 2000, A\&A, 363, 507

\bibitem[2001]{gkw01}   Gopal-Krishna, \&  Wiita, P.J. 2001,  New Astron. Rev.,
Proc.\ STScI Workshop on Lifecycles of Radio Galaxies, in press

\bibitem[1996]{gkwh}   Gopal-Krishna, Wiita, P.J., \&  Hooda, J.S. 1996, A\&A,
316, L13 (GKWH96)

\bibitem[1989]{gkws89}  Gopal-Krishna, Wiita, P.J., \& Saripalli, L. 1989, MNRAS, 239, 173
(GKWS89)

\bibitem[1998]{hardcastle} Hardcastle, M.J. 1998, MNRAS, 298, 569

\bibitem[1986]{heckman}  Heckman, T.M., et al. 1986, ApJ,  311, 526

\bibitem[1999]{hirotani}  Hirotani, K., Iguchi, S., Kimura, M., \&  Wajima, K. 1999,
PASJ, 51, 263

\bibitem[1994]{hoodaetal}  Hooda, J.S., Mangalam, A.P., \&  Wiita, P.J. 1994, ApJ,  423, 116

\bibitem[1996]{hooda96}  Hooda, J.S., \& Wiita, P.J. 1996, ApJ, 470, 211

\bibitem[1998]{hooda98}  Hooda, J.S., \& Wiita, P.J. 1998, ApJ, 493, 81

\bibitem[2000]{jeyakumar}  Jeyakumar, S.,  \& Saikia, D.J. 
2000, MNRAS,  311, 397

\bibitem[2001]{jeyakumaretal} Jeyakumar, S., Wiita, P.J., Saikia, D.J., \& Hooda, J.S.
2001, submitted to ApJ

\bibitem[1992]{jorgensen} J{\o}rgensen, I., Franx, M., \& Kj{\ae}rgaard, P.
1992, A\&AS, 95, 489

\bibitem[1997]{kaiser}  Kaiser, C.R.,  \& Alexander, P. 1997, MNRAS,  286, 215

\bibitem[1997]{kaiseretal}  Kaiser, C.R., Dennett-Thorpe, J., \&
  Alexander, P. 1997, MNRAS, 292, 723

\bibitem[1998]{kodama} Kodama, T., Arimoto, N., Barger, A.J., \& Arag{\'o}n-Salamanca, A.
1998, A\&A, 334, 99

\bibitem[1994]{komissarov}  Komissarov, S.S. 1994, MNRAS,  269, 394

\bibitem[1987]{kormendy}  Kormendy, J. 1987, in 
 Structure and Dynamics
of Elliptical Galaxies, ed.\ T.\ de Zeeuw  (Reidel, Dordrecht),   78

\bibitem[1993]{laing93}  Laing, R.A. 1993, in  
 STScI Symp. 6: Astrophysical Jets, ed.\ D.\ Burgarella, M.\ Livio, \& C.\ O'Dea 
(Cambridge Univ.\ Press, Cambridge),   95

\bibitem[1996]{laing96}   Laing, R.A. 1996,  in Energy Transport in Radio
Galaxies and Quasars, ASP Conf.\ Ser., Vol.\ 100,
ed.\ P.\ Hardee, A.H.\ Bridle, \& J. Zensus
(ASP, San Francisco),  241

\bibitem[1999]{laingetal99}   Laing, R.A., Parma, P., de Ruiter, H.R.,  \& Fanti, R. MNRAS,  1999, 306, 513

\bibitem[1996]{ledlow}   Ledlow, M.J., \&  Owen, F.N. 1996, AJ, 112, 9

\bibitem[2001]{ledlow00} Ledlow, M.J.,   Owen, F.N., \& Eilek, J.A. 2001,
 New Astron. Rev.,
Proc.\ STScI Workshop on Lifecycles of Radio Galaxies, in press

\bibitem[1999]{meier}   Meier, D.L. 1999, ApJ,  522, 753

\bibitem[1997]{meieretal}   Meier, D.L., Edgington, S., Godon, P.,  
Payne, D.G.,  \& Lind, K.R. 1997, Nature, 388, 350

\bibitem[1999]{morganti}  Morganti, R., Killeen, N.E.B., Ekers, R.D., \& Oosterloo, T.A.
1999, MNRAS, 307, 750

\bibitem[1996]{norman}  Norman, M.L. 1996, in Energy Transport in Radio
Galaxies and Quasars, ASP Conf.\ Ser., Vol.\ 100,
ed.\ P.\ Hardee, A.H.\ Bridle, \& J. Zensus
(ASP, San Francisco),   319

\bibitem[1982]{normanetal}  Norman, M.L., Winkler, K.-H.A.,
 Smarr, L., \& Smith, M.D. 1982, A\&A, 113, 285

\bibitem[1997]{odea} O'Dea, C.P., \& Baum, S.A. 1997, AJ, 113, 148

\bibitem[1993]{odonoghue} 
O'Donoghue, A.A., Eilek, J.A., \& Jones, J.M. 1993, ApJ, 408, 428

\bibitem[1989]{owen89}  Owen, F., \& Laing, R.A. 1989, MNRAS, 238, 357

\bibitem[1994]{owen94}  Owen, F., \& Ledlow, M.J. 1994, in
The First Stromlo
Symposium: The Physics of Active Galaxies, ed.\ G.V.\ Bicknell,
M.A.\ Dopita, \& P.J.\ Quinn 
(ASP, San Francisco),   319

\bibitem[1991]{owen91}  Owen, F., \& White, R.A. 1991, MNRAS, 249, 164

\bibitem[1998]{owsianik} Owsianik, I., \& Conway, J.E., 1998, A\&A, 337, 69 (OC)

\bibitem[1990]{peletier} Peletier, R.F., Davies, R.L., Illingworth,
G.D., Davis, L.E., \& Cawson, M. 1990, AJ, 100, 1091

\bibitem[1996a]{reynoldsa} Reynolds, C.S.,  Di Matteo, T., Fabian, A.C.,
Hwang, U., \& Canizares, C.R. 1996, MNRAS, 283, 111P


\bibitem[1996b]{reynoldsb}  Reynolds, C.S., Fabian, A.C., Celotti, A.,  \& Rees, M.J.
1996, MNRAS,  283, 873


\bibitem[1983]{saikia}  Saikia, D.J., Shastri, P., 
Cornwell, T.J.,  \& Banhatti, D.G.
1983, MNRAS, 203, 53P

%\bibitem[1995]{saikia} Saikia, D.J., Jeyakumar, S., Wiita, P.J., Sanghera, H.S.,
%Spencer, R.E., 1995, MNRAS, 276, 1215

\bibitem[1986]{sarazin}  Sarazin, C.L. 1986, Rev.\ Mod.\ Phys., 58, 1

\bibitem[1985]{saripalli}  Saripalli, L., \& Gopal-Krishna 1985, A\&A, 149, 205

\bibitem[1982]{scheuer}  Scheuer, P.A.G. 1982, in  IAU Symp.\ 97: Extragalactic Radio
Sources, ed.\ D.S.\ Heeschen  \& C.M.\ Wade
 (Kluwer, Dordrecht),   163

\bibitem[1982]{swarup}  Swarup, G., Sinha, R.P., \& Saikia, D.J. 
1982, MNRAS, 201, 393

\bibitem[1981]{terlevich}  Terlevich, R., Davies, R.L, 
Faber, S.M., \& Burstein, D. 1981, MNRAS, 196, 381

\bibitem[2000]{valtonen}  Valtonen, M.J., \& Hei{}n\"am\"aki, P., 2000, ApJ, 530,
107

\bibitem[2000]{wang}  Wang, Z., Wiita, P.J., \& Hooda, J.S. 2000, ApJ, 534, 201

\bibitem[1990]{wiita90}  Wiita, P.J., Rosen, A., \& Norman, M.L.
1990, ApJ, 350, 545

\bibitem[1992]{wiita92}  Wiita, P.J., \& Norman, M.L. 1992, ApJ, 385, 478

\bibitem[1999]{zhang}  Zhang, H.-M., Koide, S., \& Sakai, J.-I. 
1999, PASJ, 51, 449

\bibitem[1997]{zirbel} Zirbel, E.L. 1997, ApJ,  476, 489

\end{thebibliography}
\end{document}